\documentclass[physrev,preprintnumbers,amsmath,amssymb,letterpaper,twocolumn]{revtex4-2}

\usepackage{graphicx}
\usepackage{dcolumn}
\usepackage{bm}
\usepackage{epstopdf}
\usepackage{latexsym}
\usepackage{epsfig}
\usepackage{verbatim}
\usepackage{hyperref}
\usepackage{color}

\newcommand{\be}{\begin{equation}}
\newcommand{\ee}{\end{equation}}
\newcommand{\bea}{\begin{eqnarray}}
\newcommand{\eea}{\end{eqnarray}}

\newcommand{\req}[1]{Eq.~(\ref{#1})}

\newcommand{\db}{d_\mathrm{Be}}

\begin{document}

%\preprint{APS/123-QED}

\title{Spin-Polarized Tunneling in Critically Disordered Be-Al Bilayers} % Force line breaks with \\

\author{F.N. Womack and P.W. Adams}
\affiliation{Department of Physics and Astronomy, Louisiana State University, Baton Rouge, Louisiana 70803, USA}

\author{G. Catelani}
\affiliation{JARA Institute for Quantum Information (PGI-11), Forschungszentrum J\"ulich, 52425 J\"ulich, Germany}

\date{\today}% It is always \today, today,
 % but any date may be explicitly specified

\begin{abstract}
We report spin-polarized tunneling density of states measurements of the proximity modulated superconductor-insulator transition in ultra thin Be-Al bilayers. The bilayer samples consisted of a Be film of varying thickness, $\db=0.8-4.5\,$nm, on which a 1~nm thick capping layer of Al was deposited.  Detailed measurements of the Zeeman splitting of the BCS coherence peaks in samples with sheet resistances $R\sim h/4e^2$ revealed a super-linear Zeeman shift near the critical field. Our data suggests that critically disordered samples have a broad distribution of gap energies and that only the higher portion of the distribution survives as the Zeeman critical field is approached.  This produces a counter-intuitive field dependence in which the gap apparently {\it increases} with {\it increasing} parallel field.
\end{abstract}

\maketitle

\section{Introduction}
The disorder driven superconductor-insulator transition (SIT) has been the subject of intense investigation for more than 30 years now~\cite{Hebard,Shahar,Adams2,Steiner,Goldman1,Valles1}.  Early studies suggested a relatively simple picture of the SIT in homogeneously disordered two-dimensional (2D) systems.   As the disorder of the superconductor is increased, the underlying repulsive Coulomb correlations are enhanced until they eventually overwhelm the resident superconducting correlations.  This was believed to occur at a relatively well-defined critical disorder characterized by the quantum resistance $R_Q=h/4e^2$~\cite{Hebard, Fisher}.  However, more recent studies of the SIT have shown that the disorder-driven transition can be more complex than originally thought.  Under the proper conditions, local variations in film disorder are amplified by Coulomb interactions thereby producing regions of varying gap strength.  This leads to an intrinsic puddling of the superconducting condensate as the disorder is increased thought the SIT~\cite{Carbillet,Valles2,Valles3,Valles4,Frydman}.  In this scenario the insulating side of the SIT is fundamentally bosonic in that it consists of phase-decoupled superconducting puddles each having a finite order parameter~\cite{Valles2}.   Recent studies utilizing multiply connected geometries~\cite{Welp,Latimer} show that superconducting pair correlations exist well into the insulating phase of highly disordered, nominally homogeneous, Bi films~\cite{Valles1,Valles2,Valles3}.  This phase is commonly referred to as a Bosonic insulator.  But the details of the topological structure of the phase and its corresponding gap distribution remains unclear.  Here we present a spin-polarized tunneling study of the evolution of the superconducting gap in critically disordered Be-Al bilayers that are tuned through the SIT into a Bose insulator phase via a Zeeman field.   Parallel magnetic field is used to induce a Zeeman splitting of the BCS density of states (DoS) spectrum while minimizing the orbital field broadening of the spectral features.  As the critical field is approached from below, we find that the supra-gap spin band shifts super-linearly with increasing field.  This suggests that at low Zeeman fields the tunneling conductance measures the average of a rather broad distribution of gap energies~\cite{Trivedi}.  But as the critical field is approached from below, only the highest energy portion of the distribution remains intact thereby producing an apparent {\it increase} in average gap energy with increasing field. The resulting spectra represent a direct probe of the pairing amplitude distribution across the SIT.

Our primary goal in this study was to use the Zeeman splitting of the BCS DoS spectrum to probe the superconducting phase of quasi-homogeneously disordered films which are close to the disorder-driven zero field SIT.  In order to resolve the Zeeman splitting, the superconductor must have a low spin-orbit (SO) scattering rate.  We also require that the transition temperature of the film be well above the base temperature of our fridge.   Thin Al films meet these conditions; they have a $T_c\sim2.7\,$K and an extremely low intrinsic SO rate. Unfortunately, however, Al forms granular films, even when they are quench-condensed at 84~K.  The granularity makes high resistance Al films somewhat unstable in air.  Furthermore, the transport properties of high resistance Al films can be completely dominated by the intra-grain coupling and {\it not} necessarily by many-body effects.  In order to circumvent this limitation, we deposit the Al on a thin layer of Be.   Beryllium forms dense, adherent, amorphous films on glass substrates and, like Al, it has a low SO rate \cite{Adams2}. However, the transition temperature of Be films, $T_c\sim0.5\,$K, is too low for our purposes \cite{Be2}.  The bilayer arrangement allows us to partially mitigate the granularity of the Al by providing an underlying metallic coupling between the Al grains while still maintaining a reasonably high $T_c$ by virtue of the proximity effect~\cite{PE}.

\section{Experimental methods}
Superconducting Be-Al bilayer films were formed by first depositing a thin Be layer of varying thickness onto a fire-polished glass substrate followed by the deposition of a 1 nm-thick Al film.  The depositions were performed by e-beam evaporation from 99.9\% Be and 99.999\% Al targets at a rate of $\sim0.2\,$nm/s.  The glass substrates were maintained at 84~K during the deposition of both the Be and Al layers.  The bilayers were deposited without breaking a vacuum of $P<3\times10^{-7}$ Torr.  Planar tunnel junctions were formed between the upper Al layer of the samples and a counter-electrode composed of a non-superconducting Al alloy using a 1 nm layer of SiO as the tunnel barrier.  Bilayers with an Al thickness of 1~nm and Be thicknesses ranging from 0.8 to 4.5~nm had normal state sheet resistances that ranged from $R\approx100\,\Omega$ to $10^4\,\Omega$ at low temperature. Magnetotransport measurements were made on a Quantum Design Physical Properties Measurement System He3 probe.  The maximum applied field was 9~T and the base temperature of the system was 400~mK. The tunneling measurements were carried out using a standard 27~Hz 4-wire lock-in amplifier technique.
\section{Results and discussion}

In general the critical field of a thin film superconductor has both an orbital and a Zeeman component~\cite{Fulde}. If one makes a low atomic mass film~\cite{SO}, such as Al, sufficiently thin and orients the field parallel to the film surface then the orbital response will be suppressed and one can realize a purely Zeeman-mediated critical field transition~\cite{Adams3,Be2}, which is first order at low temperatures. The $T=0$ parallel critical field is given  by the Clogston-Chandrasekhar equation~\cite{CC}, $H_{c\parallel}=\frac{\sqrt2\Delta_0}{g\mu_{\rm B}}$, where $\Delta_0$ is the zero temperature - zero field gap energy, $\mu_{\rm B}$ is the Bohr magneton, and $g$ is the Land\'{e} g-factor.  In this series of experiments we have explored the Zeeman response of Be/Al bilayers with sheet resistances ranging from $R\ll R_Q$ to $R\sim R_Q$.

\begin{figure}
\begin{flushleft}
\includegraphics[width=.44\textwidth]{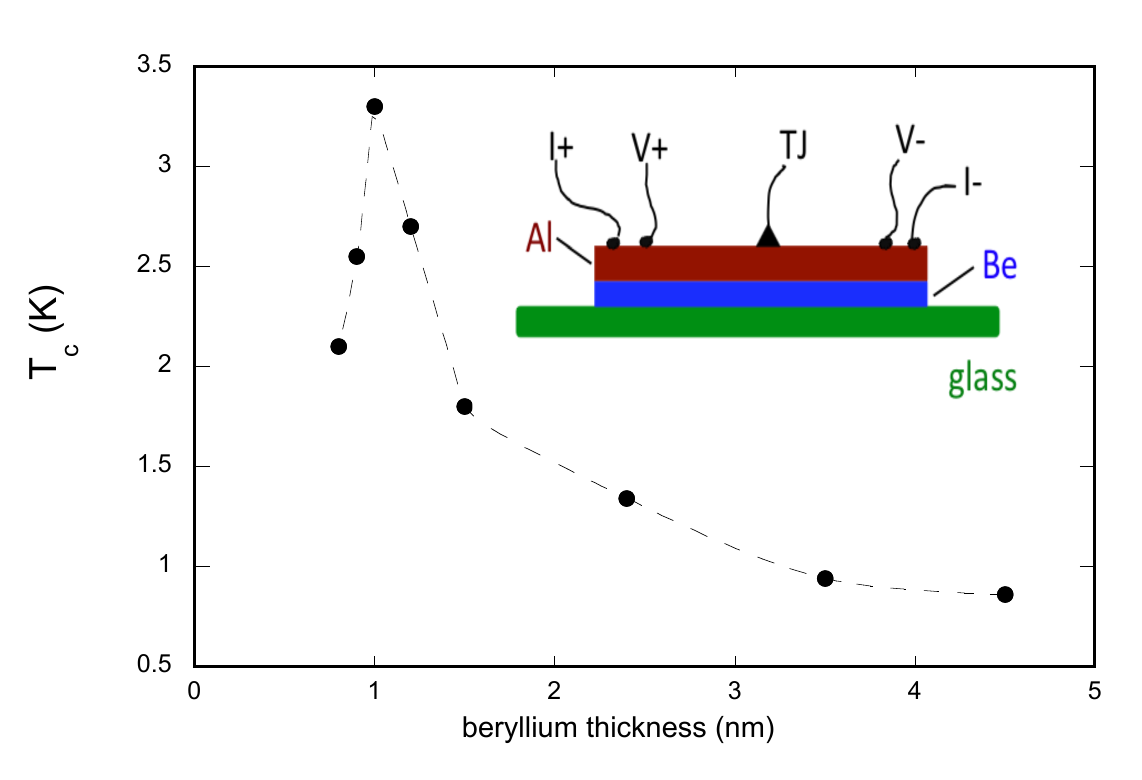}\end{flushleft}
\caption{Plot of the bilayer transition temperature as a function of the Be layer thickness $\db$.  Inset: schematic diagram of the sample geometry.  The thickness of the Be layer was systematically varied between 0.8~nm and 4.5~nm, while the Al layer thickness was maintained at 1 nm.}
\label{Tc-t}
\end{figure}

Shown in Fig.~\ref{Tc-t} are the bilayer transition temperatures $T_c$ as a function of the Be thickness.  For Be thicknesses greater than $\db\sim1.5\,$nm the bilayer transition temperatures lie between that of pure Be film, $T_c^{\rm Be}\sim0.5\,$K, and that of a pure 2.5 nm Al film, $\sim2.7\,$K.  The fact the $T_c$ decreases with increasing $\db$ in this range suggests that the bilayer transition temperature was mediated by the proximity effect \cite{PE}.  In the thickness range $\db<1.5\,$nm, there is a local maximum in $T_c$.   In the limit $\db\rightarrow0$, $T_c$ should be asymptotic to the transition temperature of the underlying Al layer.  Unfortunately, the $T_c$ of the 1 nm Al layer is unknown because the bilayers become electrically discontinuous for $\db\lesssim0.5\,$nm.  Nevertheless, it is possible that the Al layer has a transition temperature well above that of a 2.5 nm Al film. In fact, the data in Fig.~\ref{Tc-t} suggests $T_c^{\rm Al}\sim3.5\,$K. Bilayers with $\db\lesssim1$ nm have sheet resistances approaching the quantum resistance $R_Q=h/4e^2$ \cite{Fisher}.  Since $R_Q$ represents the threshold for the SIT we believe that the local maximum in Fig.~\ref{Tc-t} arises from the preemptive effects of increasing disorder.

\begin{figure}
\begin{flushleft}
\includegraphics[width=.44\textwidth]{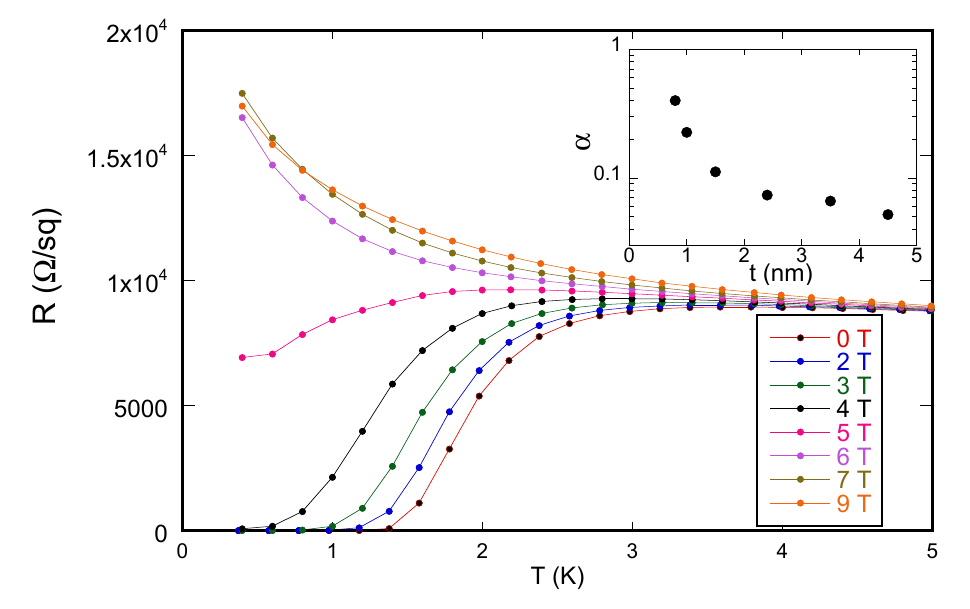}\end{flushleft}
\caption{Superconductor-insulator transition driven by a parallel magnetic field for an Be-Al bilayer with a beryllium thickness of 0.8~nm and an aluminum thickness of 1~nm.   Inset: Relative depth of the zero bias tunneling anomaly $\alpha$ (defined in the text) as a function of beryllium layer thickness.}
\label{R-T-H}
\end{figure}

In Fig.~\ref{R-T-H} we plot the temperature dependence of the sheet resistance of a $\db=0.8\,$nm bilayer in a range of applied parallel magnetic fields.  The ``fan-like'' structure of the data is associated with the field-tuned SIT and has been reported in a wide variety of critically disordered thin film superconductors \cite{Hebard,Goldman1,Shahar,Valles4}.  Note the separatrix between the superconducting and insulating phases at a transport parallel critical field $H_{c\parallel}^{\mathrm{tr}}\sim5\,$T.  These data demonstrate that our thinnest Be layer samples are near the zero field SIT and that their underlying normal state is insulating \cite{Lee,AAGS}.  Furthermore, there is an obvious low temperature crossing of the 7 T and 9 T traces.  This suggests that there is a possible superconducting contribution to the insulating behavior over a narrow range of fields.  In other words, the data in Fig.~\ref{R-T-H} is consistent with the emergence of a Bose insulator phase at intermediate parallel fields.   Below we present a tunneling density of states study of the Zeeman-tuned SIT via the application of parallel magnetic~\cite{Catelani3}.

\begin{figure}
\begin{flushleft}
\includegraphics[width=.44\textwidth]{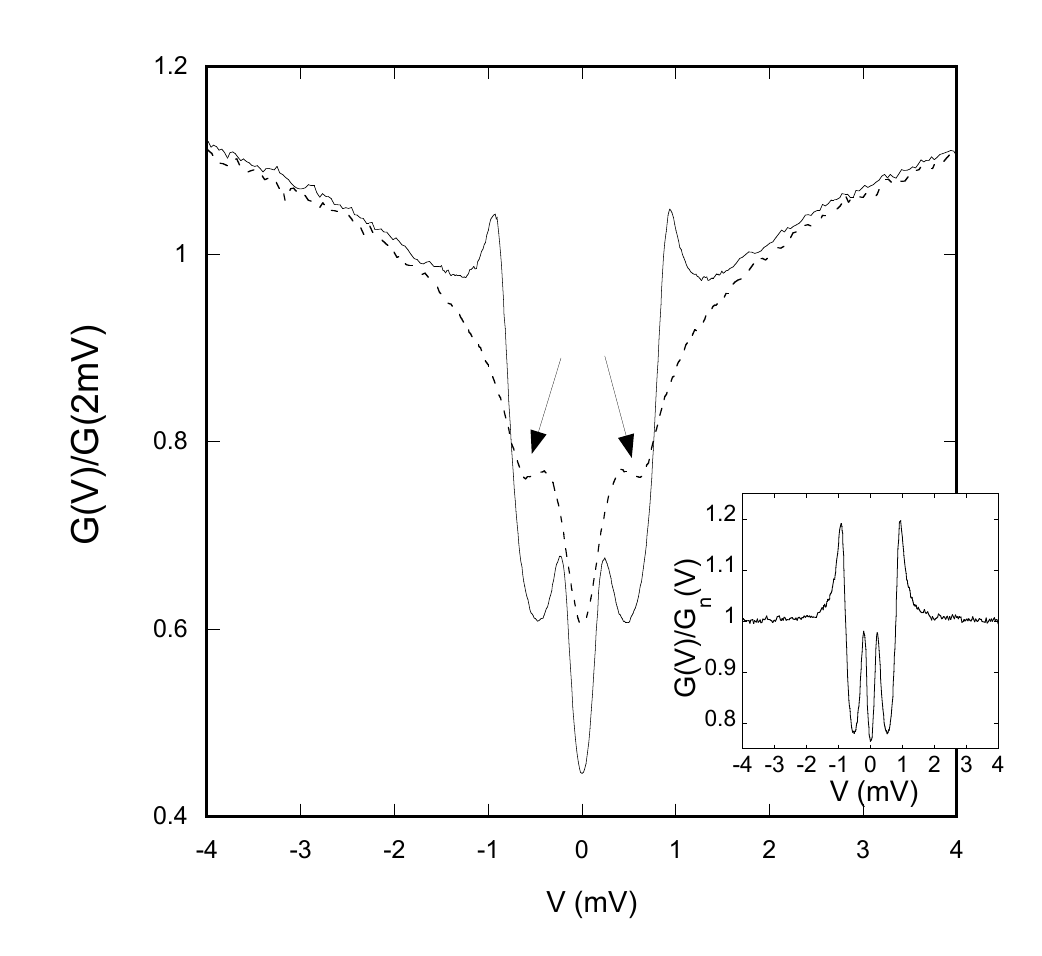}\end{flushleft}
\caption{Tunneling conductance as a function of bias voltage for the bilayer used in Fig.~\ref{R-T-H} taken at $T=400\,$mK.  The data are normalized by the conductance at 2 mV relative to the Fermi energy.  The solid line represents the Zeeman-split BCS density of states in the superconducting phase of the bilayer in a parallel field of $H_\parallel=6$ T.  The dashed line represents the normal state spectrum in a supercritical field $H_{\parallel}=7$ T.  Note the substantial suppression of normal state tunneling conductance near $V=0$.  The arrows point to the pairing resonance features in the normal state spectrum.  Inset: normalized superconducting density of states after the normal state spectrum obtained in a 9 T {\it perpendicular} field was divided out of the raw superconducting data.}
\label{Spectra}
\end{figure}

Shown in Fig.~\ref{Spectra} are tunneling conductance spectra taken on the bilayer of Fig.~\ref{R-T-H}.  At low temperatures the tunneling conductance $G$ is simply proportional to the single-particle density of states (DoS)~\cite{Tinkham}.  The bias voltage is relative to the Fermi energy and the conductances have been normalized by the conductance at 2 mV.  The solid trace in the upper panel represents the tunneling conductance in the superconducting phase of the bilayer and the dashed trace is the corresponding conductance in the high-field normal phase. We note that based on the DoS measuerements, the parallel critical field, $H_{c\parallel}\sim 7\,$ T, is higher than that estimated from transport measurements; we will return on this point later. The are three features of the tunneling spectra in Fig.~\ref{Spectra} that are of particular importance to this study.  The first is the Zeeman splitting of the BCS coherence peaks \cite{Fulde,Meservey,Adams4}.  The second is a broad logarithmic suppression of the DoS in the normal phase of the bilayer.  This suppression arises from $e-e$ interaction effects~\cite{AAGS} and is often referred to as the zero bias anomaly (ZBA) or the Coulomb anomaly~\cite{Adams2}.  It is a direct microscopic measure of the disorder-induced repulsive correlations and has been well-documented in a wide variety of 2D systems.  The third is the pairing resonance (PR) represented by the dips riding on top of the normal state spectra~\cite{Adams6,Altshuler2}, as indicated by the arrows in Fig.~\ref{Spectra}.  Details of the PR have been published elsewhere~\cite{Catelani}, but for our purposes the resonance, which arises from evanescent Cooper pairs, provides us with a direct probe of the spin properties of the normal state (see Appendix~\ref{app:PR}). In particular, we can use the PR to extract the effective normal state g-factor, which may differ from the naive value $g=2$ due to Fermi liquid (FL) effects~\cite{Catelani2}.

As the beryllium thickness is decreased below 1~nm both the resistance of the bilayers and the magnitude of the ZBA increase precipitously.  This is due, in part, to the fact that a 1~nm Al film deposited directly on the glass substrate would not be electrically continuous. Therefore, for our chosen geometry the Be thickness controls the level of disorder.  In order to quantify the strength of the ZBA we define the dimensionless parameter $\alpha=[G(2mV)-G(0)]/G(2mV)$ which measures the relative depletion of electron states at the Fermi energy in the high-field normal phase.   A plot of $\alpha$ as function of beryllium thickness is shown in the inset of Fig.~\ref{R-T-H}.  Note that $\alpha$ grows rapidly as $\db$ is lowered below 2~nm, indicative of the approach to the zero-field SIT.   As is evident in Fig.~\ref{Spectra} the depletion of single particle states due to the ZBA and the depletion due to opening of the superconducting gap are comparable in magnitude in our most disordered samples.

\begin{figure}
\begin{flushleft}
\includegraphics[width=.44\textwidth]{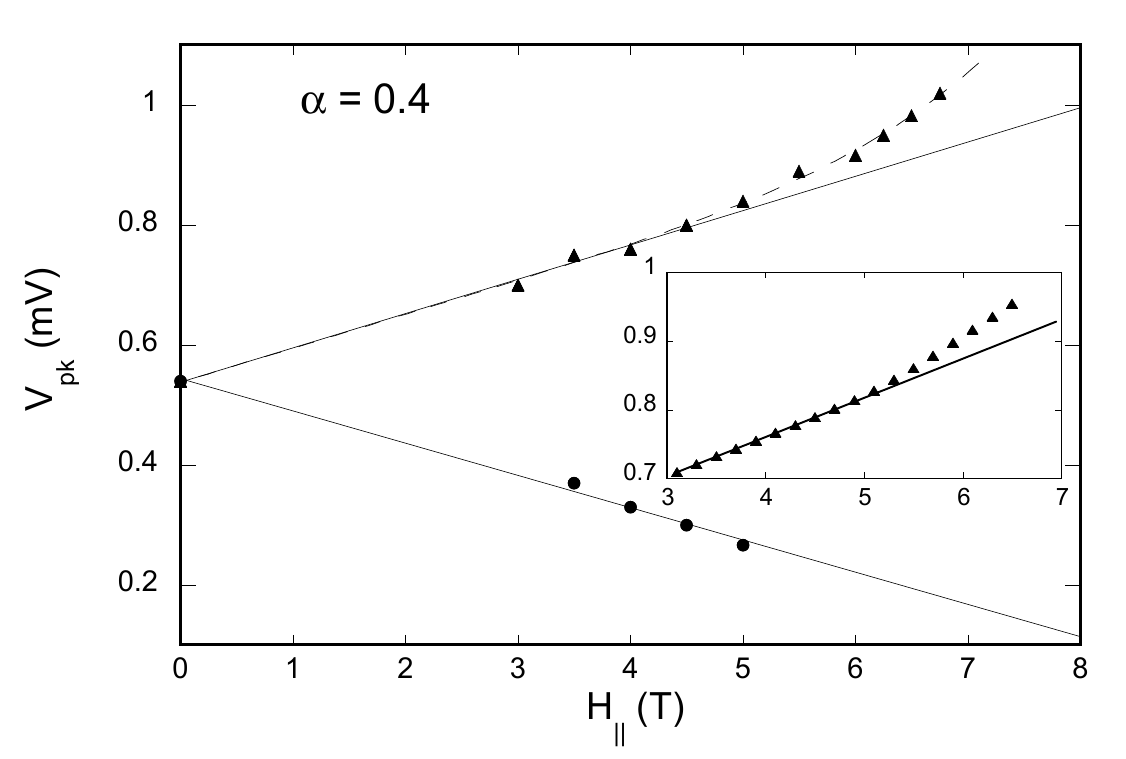}\end{flushleft}
\caption{Position of the supra-gap (triangles) and sub-gap (circles) coherence peaks as a function of parallel field for a bilayer with $d_{\rm Be}=0.8\,$nm.  The solid lines represent the expected Zeeman splitting, the dashed line is a guide for the eye. Inset: supra-gap peak position (thick line) from the DoS averaged over a Gaussian distribution of zero-field gaps centered around $\bar{\Delta}$. The line is the expected linear Zeeman splitting. The value of $\bar{\Delta}$ has been chosen to match the extrapolated zero-field peak position of the experimental data; this also fixes the field scale as $\bar{\Delta}/\mu_B$.}
\label{PeakPositions}
\end{figure}

Shown in the lower panel of Fig.~\ref{Spectra} is the superconducting spectrum of the upper panel after the ZBA spectrum has been divided out of the data.  In order to suppress the PR the normal state ZBA spectrum was measured by applying a 9 T {\it perpendicular} field.  The four peaks, the two outer supra-gap peaks and the two inner sub-gap peaks, represent the Zeeman splitting of the BCS spin-up and spin-down subbands \cite{Fulde,Meservey,Adams4}.  The occupied and unoccupied subband peaks are located at $eV=\pm(\Delta_0\pm \frac{1}{2}eV_Z)$ where $eV_Z=g\mu_{\rm B}H_\parallel$ is the Zeeman energy.  Although these data were taken relatively close to the parallel critical field $H_{c\parallel}\sim7$ T, the peaks are sharp and their positions can, in principle, be measured quite accurately through the parallel critical field transition.  In Fig.~\ref{PeakPositions} we plot the position of the supra-gap and sub-gap coherence peaks as a function of parallel field for a $\db=0.8\,$nm bilayer having an $\alpha\sim0.4$.  However, near $H_{c\parallel}$ the spectrum begins to display characteristics of both the superconducting and normal phases.   In particular, the PR dip emerges and is superimposed on the superconducting spectrum.  The position of the PR,
\be
eV^*=\frac{1}{2}\left[eV_Z + \sqrt{(eV_Z)^2-\Delta_0^2}\right]
 \label{PR}
\ee
is quite close to, and partially overlaps with, the superconducting sub-gap peaks near $H_{c\parallel}$.  This makes it difficult to determine the sup-gap positions in the critical region. The supra-gap peaks, however, are positioned well away from the PR, so we have chosen to focus our analysis on their field dependence.

\begin{figure}
\begin{flushleft}
\includegraphics[width=.44\textwidth]{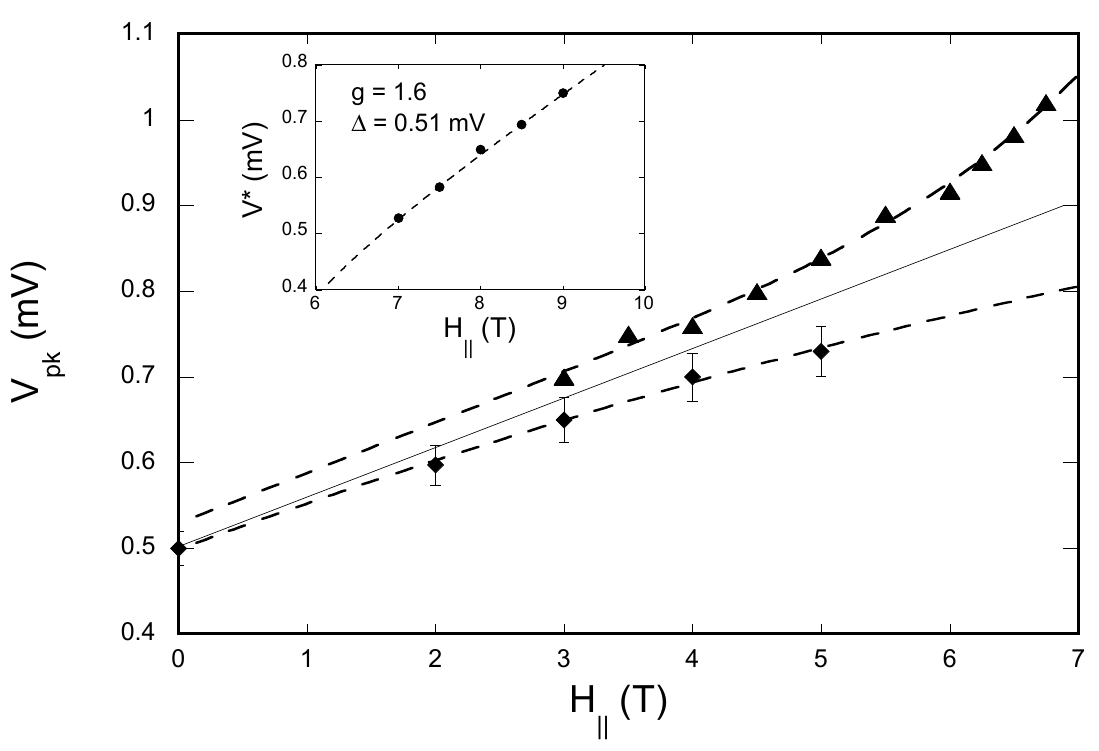}
\end{flushleft}
\caption{ Position of the supra-gap coherence peak for the bilayer of Figs.\ref{R-T-H} and \ref{PeakPositions} (diamonds) and a low resistance pure 2.5~nm thick Al film (circles). The dashed lines are provided as a guide to the eye.  The solid line represents the expected linear Zeeman dependence.  Inset: position of the normal state pairing resonance of the bilayer.  The dashed line is a least-squares fit to \req{PR}}
\label{Vup}
\end{figure}

The supra-gap peak positions from the bilayer of Fig.~\ref{PeakPositions} are shown in Fig.~\ref{Vup} along with the corresponding peaks in a moderately disordered Al film, $R\sim1000~\Omega$.  The solid line in this plot represents $V_Z$ with $g=2$.  At fields well below $H_{c\parallel}$ the supra-gap peaks in both films exhibit the expected Zeeman shift.  However, as the critical field is approached the Al data falls below the Zeeman line while the Be-Al data rises super-linearly.   The sub-linear field dependence of the Al peak is due to both a downward FL renormalization of the quasiparticle spin~\cite{Catelani2} and to small, but finite, orbital pair-breaking effects.  The former can be estimated by measuring the PR position as a function of field, see inset of Fig.~\ref{Vup}.  The dashed line in the inset represents a least-squares fit of \req{PR} to the data in which $g$ and $\Delta_0$ where varied. The fit clearly shows that the normal state g-factor in the bilayers is also well below 2.  Therefore, the super-linear field dependence of $V_{\rm pk}$ cannot be due to FL exchange effects.

The super-linear field dependence of $V_{\rm pk}$ was only observed in highly disordered bilayers having $\alpha\gtrsim0.2$ (see Appendix~\ref{app:1nm}), which suggests that the effect is a consequence of disorder-enhanced $e-e$ interactions as manifest in the ZBA.  In the strong disorder limit the system begins to break up into weakly connected superconducting islands which have a broad range of local gap energies \cite{Trivedi,dlo,Roy}.  Thus, the super-linear field dependence of $V_{\rm pk}$ reflects the portion of the gap distribution which survives as $H\rightarrow H_{c\parallel}$.  At low fields the tunneling conductance samples the entire distribution of gaps but near the critical field the sampling is skewed towards high gap values.  We have modeled this behavior by averaging over a Gaussian distribution of zero-field gaps (see Appendix~\ref{app:Gauss}). This approach is oversimplified at low field, where different parts of the sample are all superconducting and can influence each other directly, but should be qualitatively correct at high fields, where the surviving superconducting regions are disconnected~\cite{dlo} (a fully self-consistent calculation as those presented in Refs.~\cite{Trivedi,dlo} is beyond our scope). The result of such a calculation displays a supra-linear behavior resembling the experimental data, see the inset in Fig.~\ref{PeakPositions}. The existence of isolated superconducting regions between 5 and 7~T could also explain the discrepancy between transport and DoS estimates for the parallel critical field. Interestingly, fitting the DoS with a phenomenological Dynes broadening, while inadequate, also supports our interpretation, see Appendix~\ref{app:Dynes}.

\section{Summary}
In summary, we have exploited the Zeeman splitting of the BCS DoS spectrum  to obtain direct evidence for a distribution of pairing energies in a critically disordered BCS superconductor.  The effects of disorder serve to both suppress $T_c$ and broaden the critical field transition.  Our data suggest that the Zeeman-tuned SIT is dominated by the tail of a broad distribution of local gap energies.  Presumably this high energy tail also plays a prominent role in the zero-field, disorder-driven SIT transition.

\acknowledgments
The magneto-transport measurements were performed by P.W.A. and F.N.W with the support of the U.S. Department of Energy, Office of Science, Basic Energy Sciences, under Award No.\ DE-FG02-07ER46420.    The theoretical analysis was carried out by G.C.

\appendix

\section{Pairing resonance}
\label{app:PR}

\begin{figure}[tb]
\begin{flushleft}
\includegraphics[width=.44\textwidth]{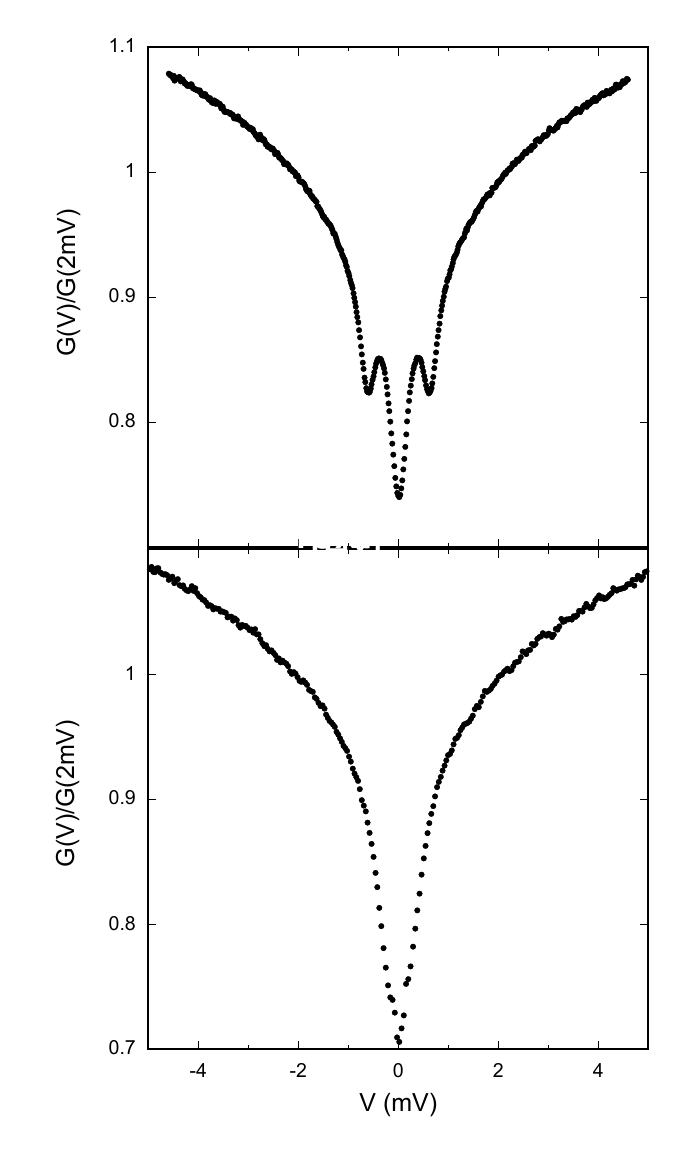}\end{flushleft}
\caption{High field tunneling conductance as a function of bias voltage at 400 mK.  The upper panel is the tunneling conductance of the $\db=1.0$ nm bilayer of Figs.\ \ref{R-T-H_2} and \ref{G(V)} in a parallel field of 9 T.  The lower panel is the tunneling conductance in the same bilayer in a perpendicular field of 4.0 T.  Note that the pairing resonance satellite features are completely extinguished in the perpendicular field. }
\label{BG}
\end{figure}
A low spin-orbit BCS superconductor in the presence of a pure Zeeman field will undergo a low-temperature first order transition to the normal state at a critical field that is simply proportional to the gap energy~\cite{Fulde},
\be
H_{c\parallel}=\frac{\sqrt2\Delta_0}{g\mu_{\rm B}}.
\ee
For fields above $H_{c\parallel}$ the system enters a paramagnetic phase and the mean-field BCS order parameter is quenched.   Although there is no global superconducting order parameter in the paramagnetic phase, a well-pronounced superconducting fluctuation mode still exists \cite{KAA}. Electrons that tunnel with the proper energy can, in fact, produce a resonant excitation of this mode and thereby cause a strong tunneling DOS singularity which we term the pairing resonance (PR). The bias voltage $V^*$ corresponding to the resonant mode and the corresponding DOS singularity is a function of the supercritical field~\cite{KAA}, see Eq.~(\ref{PR}).
%\be
%eV^*=\frac{1}{2}\left[eV_Z + \sqrt{(eV_Z)^2-\Delta_0^2}\right]
% \label{PR}
%\ee
%where $eV_Z=g\mu_{\rm B}H_{\parallel}$ is the Zeeman energy.  
The mode persists to fields well beyond $H_{c\parallel}$.  As the field increases the PR  broadens and moves out to higher energy as per \req{PR}.  Note that the implicit g-factor $g$ dependence in \req{PR} can be used to determine $g$ in the paramagnetic phase by measuring $V^*(H_\parallel)$ and the then fitting the data to \req{PR}.  The PR is extremely sensitive to both spin-orbit scattering and perpendicular field.  Both of these strongly suppress the resonance.  Shown in the upper panel of Fig.~\ref{BG} is the normal state tunneling spectrum for the $\db=1.0$ nm bilayer in a 9 T parallel field.  The satellite dips are the PR which are superimposed on the $e-e$ interaction mediated ZBA \cite{AAGS}.  The lower panel is the resulting spectrum when a 4 T {\it perpendicular} field is applied.  Note that PR is completely suppressed but the background ZBA remains largely unaffected.

\section{Parallel-Field SIT in a 1 nm - 1 nm Be-Al bilayer}

\label{app:1nm}

\subsection{Transport}

In Fig.~\ref{R-T-H_2} we plot the temperature dependence of the sheet resistance of a bilayer having layer thicknesses $\db=1.0\,$nm and $d_{\rm Al}=1.0\,$nm in a range of applied parallel magnetic fields.  The ``fan-like'' structure of the data is very similar to that of the higher resistance $\db=0.8\,$nm bilayer shown in Fig.~\ref{R-T-H}.   The transport data suggests a that the $T=0$ SIT critical field for this sample is $H_{c\parallel}=7\,$T.  Note that the 7 T traces separates the superconducting-like and insulating-like temperature dependencies in Fig.~\ref{R-T-H_2}.   Indeed the 7~T trace suggests that an intermediate metallic phase exists at the boundary of the SIT. The insulating behavior of this bilayer is not as pronounced as that of the one in Fig.~\ref{R-T-H}, but this is expected given that the normal state resistance of the former is less than half that of the latter.  Interestingly, the crossing of the 8~T and 9~T traces in Fig.~\ref{R-T-H_2} is similar to what is seen in Fig.~\ref{R-T-H}.

\subsection{Tunneling}

\begin{figure}[tb]
\begin{center}
\includegraphics[width=.4\textwidth]{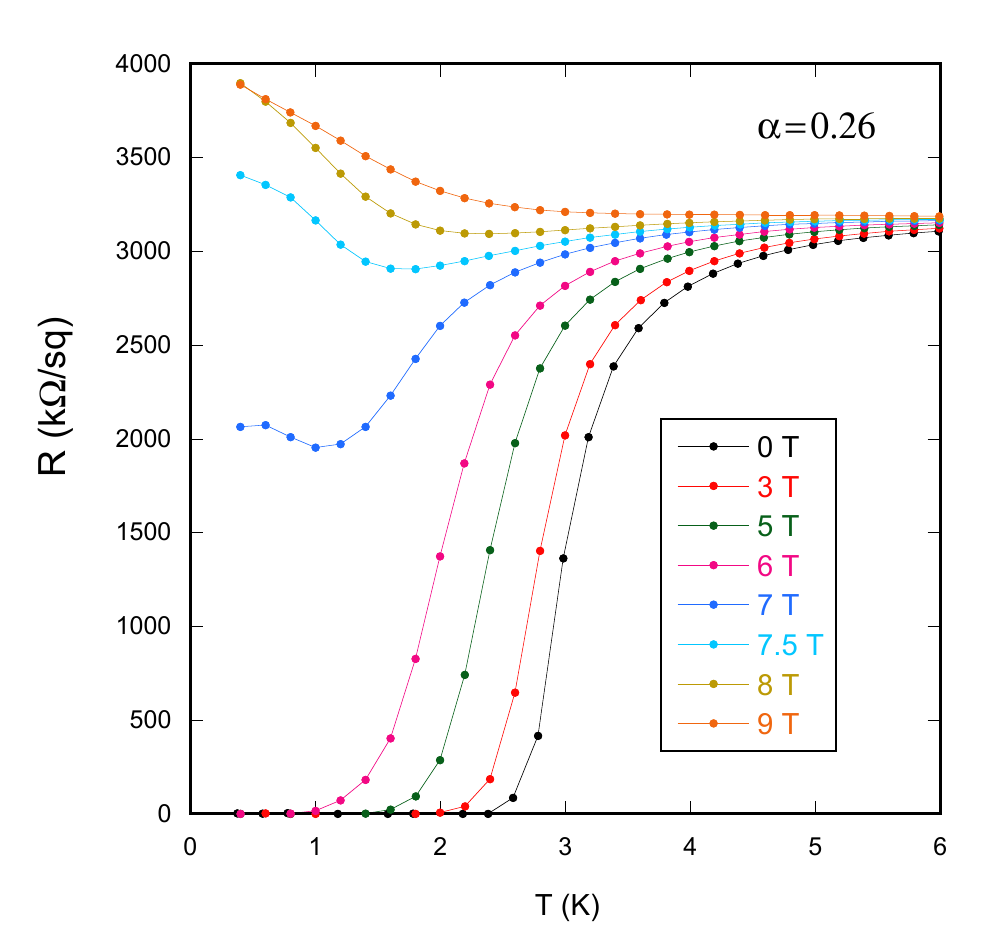}\end{center}
\caption{Superconductor-insulator transition driven by a parallel magnetic field for an Be-Al bilayer with a beryllium thickness $d_{\rm Be}=1.0$ nm and an aluminum thickness of 1 nm.   This bilayer had a lower resistance and a smaller tunneling anomaly depth ($\alpha=0.26$) than that of the $d_{\rm Be}=0.8\,$nm bilayer presented in the main text.}
\label{R-T-H_2}
\end{figure}

\begin{figure}[tb]
\includegraphics[width=.45\textwidth]{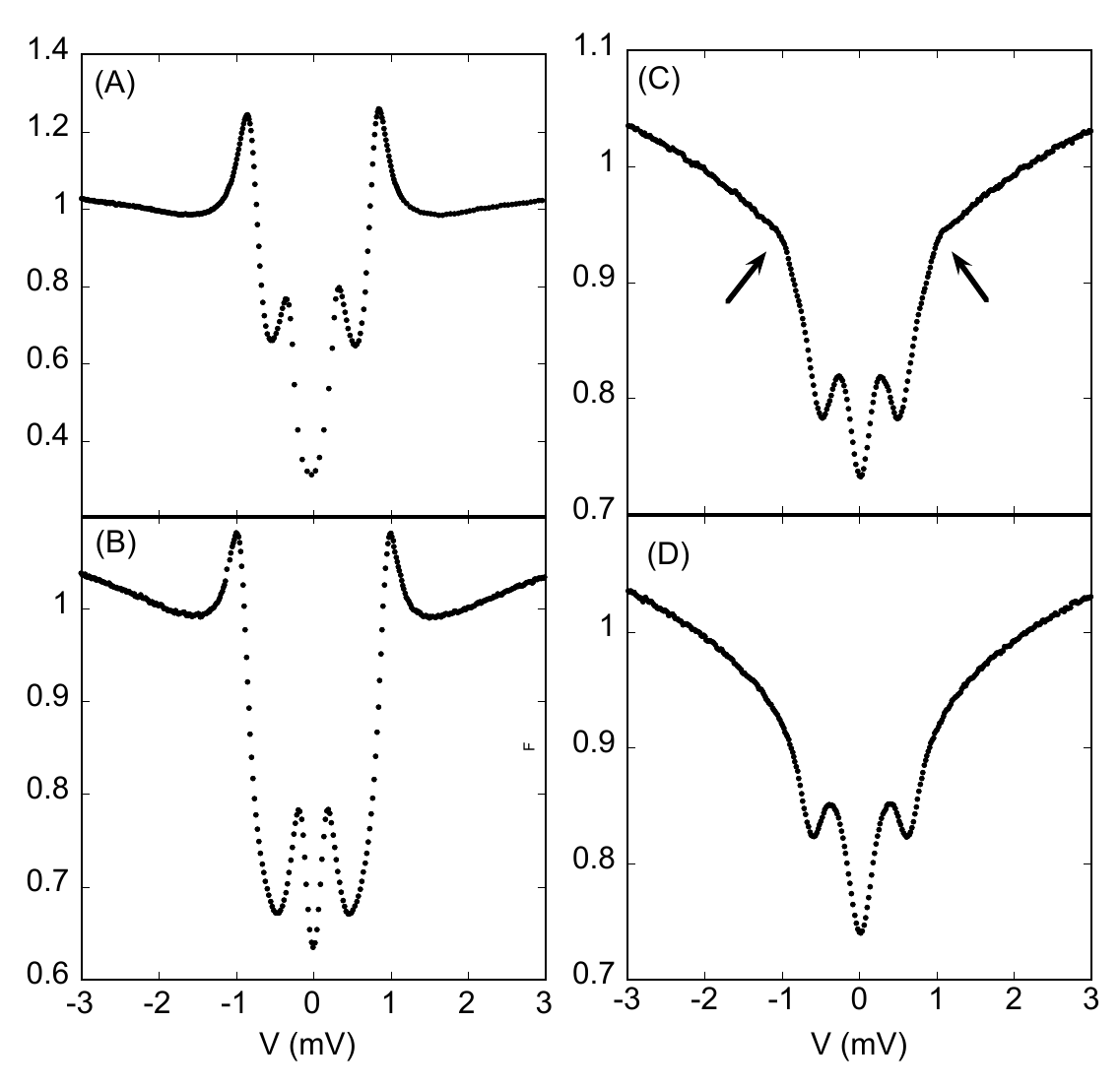}
\caption{Tunneling conductance for the bilayer shown in Fig.~\ref{R-T-H_2} taken at 400 mK.  Spectra are presented at several different parallel fields spanning the SIT.   The y-axes represent the tunneling conductance $G(V)$ normalized by its value at $V=2\,$mV and the x-axes are the tunnel junction bias voltage. (A) Spectrum from deep in the superconducting phase $H_\parallel=4.5\,$T. (B) Spectrum in what appears to be an intermediate metallic phase, $H_\parallel=7.0\,$T.  (C) Spectrum just on the insulating side of the SIT, $H_\parallel=7.5\,$T. (D) Normal state spectrum showing the pairing resonance dips on either side of the ZBA, $H_\parallel=9.0\,$T}
\label{G(V)}
\end{figure}

Here we present tunneling density of states measurements across the SIT shown in Fig.~\ref{R-T-H_2}.  Tunneling allows us to establish the relationship between the measured gap and the transport characteristics at a given parallel field.  In panel (A) of Fig.~\ref{G(V)} we plot the normalized tunnel conductance as a function of bias resistance at $H_{\parallel}=4.5\,$T.  As can be seen in Fig.~\ref{R-T-H_2} the system is well in the superconducting phase at this field which is reflected in the corresponding tunneling spectrum.  The classic BCS coherence peaks are clearly evident but have been Zeeman-split into spin-up and spin-down sub-bands by the applied field.  Interestingly, in panel (B) we see a similar superconducting spectrum at a substantially higher field of $H_{\parallel}=7.0\,$T although the corresponding 7.0~T transport trace in Fig.~\ref{R-T-H_2} suggests that the system is in a metallic phase.  At $7.5\,$T the transport clearly exhibits insulating behavior but the corresponding tunneling spectra in panel (C) of Fig.~\ref{G(V)} indicates that a finite superconducting condensate persists.  Although the coherence peaks have been suppressed in this spectrum, a sharp break in the background (indicated by the arrows) marks superconducting contribution to the tunneling conductance.  This incommensurability between transport and tunneling is similar to what we observed in the $\db=0.8\,$nm bilayer in the main text.  Specifically, these data suggest that local superconductivity plays a role the insulating behavior observed in the transport.  Finally in panel (D) we show what we believe to be a normal state spectrum at 9.0 T.  All superconducting signatures are absent save for the pairing resonance, which appears as satellite dips superimposed on the zero bias anomaly (ZBA) background \cite{AAGS}.  Note that the corresponding 9.0~T transport trace in Fig.~\ref{R-T-H_2} appears to be less insulating than that of the 8~T suggesting that there may be a bosonic contribution to the insulating behavior in the field range of 7 to 9~T.
	
\subsection{Supra-gap peak position}

\begin{figure}[tb]
\begin{flushleft}
\includegraphics[width=.44\textwidth]{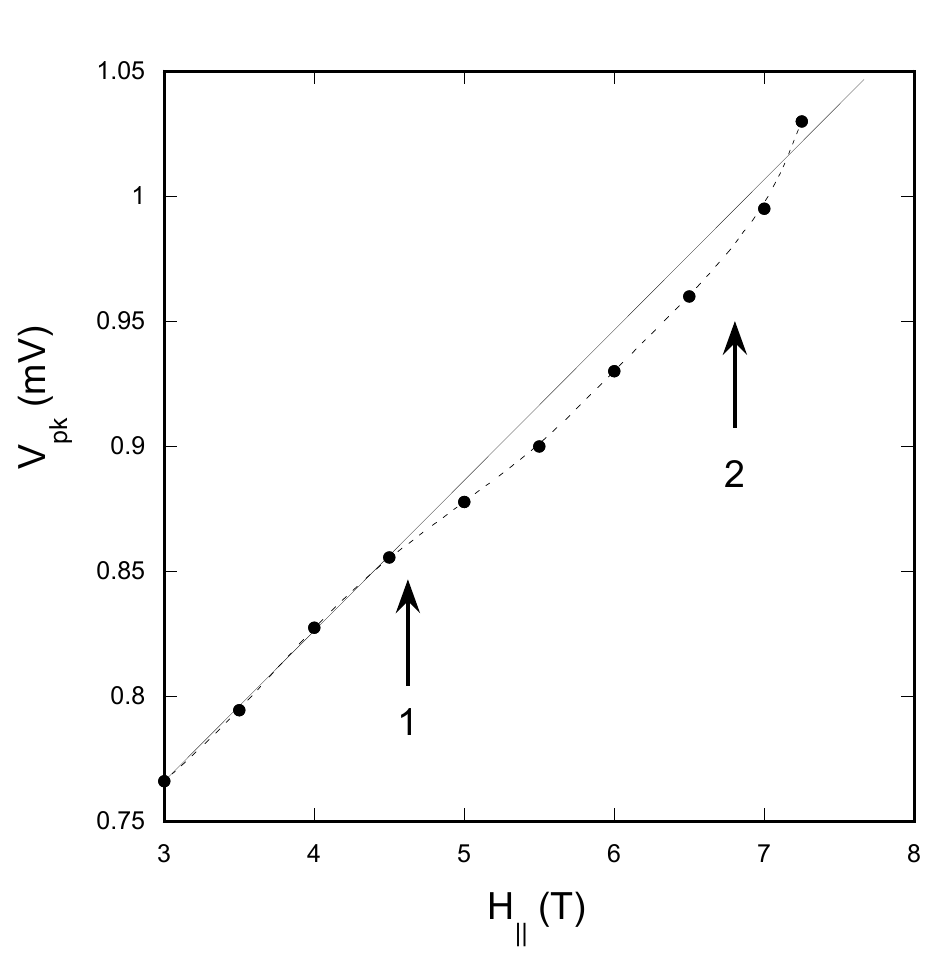}\end{flushleft}
\caption{Supra-gap peak position as a function of parallel field from tunneling spectra on the bilayer of Fig.~\ref{G(V)}.  The solid line represents the expected linear Zeeman field dependence.  The arrows indicate where (1) the measured peak positions begin the fall below the Zeeman line and (2) where there field dependences becomes super-linear.}
\label{Vup2}
\end{figure}

In Fig.~\ref{Vup2} we plot the supra-gap peak position as a function of parallel field as obtained from the spectra represented in Fig.~\ref{G(V)}.  This plot corresponds to Fig.~\ref{Vup}.  The solid line represents the linear Zeeman dependence assuming $g=2$.  Typically the supra-gap peak positions in a relatively low disorder film will fall sightly below the Zeeman line, see Fig.~\ref{Vup}.   This is primarily due to the fact that the parallel field causes a small suppression of the mean-field gap via orbital pair breaking.  Although the bilayer thickness is much less than the coherence length, the orbital suppression of the gap cannot be completely attenuated.  Furthermore, the g-factor in the normal state is somewhat less than 2 due to Fermi liquid exchange effects and as the critical field is approached the quasiparticle exchange turns on thereby effectively decreasing $g:2\rightarrow1.6$ \cite{Catelani}.   Arrow 1 in Fig.~\ref{Vup2} marks the initial break from the Zeeman line that is later followed by the emergence of super-linear field dependence in the region of arrow 2.   We believe that the this super-linear field dependence arises from a distribution of gap energies as was the case for Fig.~\ref{Vup}.  However, since this bilayer is substantially less disordered the effect is somewhat less pronounced but nevertheless clearly evident.

\section{Gaussian model of independent superconducting gaps}
\label{app:Gauss}

Experimental evidence for a roughly Gaussian distribution of local superconducting gaps has been collected over the years in disparate superconducting material such as high-$T_c$ superconductors~\cite{davis}, iron-based superconductors~\cite{singh}, and low-$T_c$ materials~\cite{carbillet}. The gap distribution standard deviation can be a significant fraction of the average gap: 20~\%, 14~\%, and 6~\%, respectively, in the cited works. Since the critical fields increase with the gap, the local gap variation can lead to a local variation of the critical fields. In an extremely simplified model, we treat different parts of a sample as separate and independent superconducting puddles. We do not expect such model to be realistic at low field, where all the puddles are superconducting and therefore can influence each other via the proximity effect; however, if at high field a sufficient portion of the puddles have turned normal, such a treatment should be qualitatively correct.

\begin{figure}[!tb]
\includegraphics[width=0.47\textwidth]{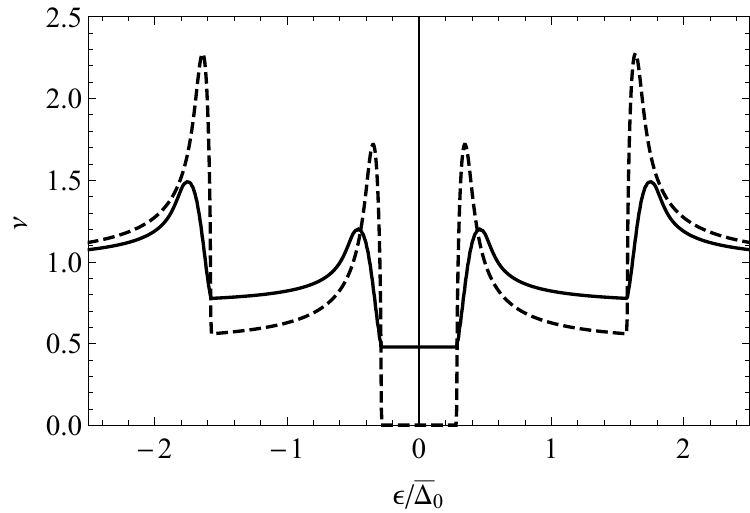}
\caption{Density of states averaged over a Gaussian gap distribution (solid line) and for a single value of the gap (dashed) -- see text for more details.}
\label{fig:dos}
\end{figure}

In our numerical calculations, we proceed as follows: we start with a Gaussian distribution of gaps with the standard deviation $\sigma$ being 10~\% of the average zero field, zero temperature gap value $\bar{\Delta}_0$ and we truncate the distribution to $\pm2\sigma$. We then divide the distribution into 39 bins and normalize the weights so that the sum of all weights is 1. From the zero-field, zero-temperature gap value $\Delta_0$ of each bin, we self-consistently calculate the order parameter as function of field, taking into account the orbital effect of the parallel field $H$ as well as the temperature (0.4~K) at which the measurement is done. The approach used here is a simplified version of that in Ref.~\cite{Catelani2}, since we are neglecting the possible effects of spin-orbit scattering and Fermi-liquid renormalization of the spin susceptibility. For the orbital parameter $c=0.02$, we have used the value extrapolated from measurements in thicker Al films.

Once the dependence of the order parameter on parallel field is known, we can calculate the energy- and field-dependent superconducting density of state (DoS) $\nu_{\Delta_0} (\epsilon, H)$ for each gap value $\Delta_0$; finally, for a given field $H$ we average over the Gaussian distribution by performing a weighted sum of the densities of states. If $H$ exceed the maximum possible field~\cite{shf} for a given $\Delta_0$, the corresponding DoS is taken as normal-metal (i.e., energy independent) one. As an example, we show in Fig.~\ref{fig:dos} a plot of the \textit{average} DoS calculated at the maximum field for the average gap (solid line) and for comparison the DoS calculated at the same field for a \textit{single} value of the gap corresponding to the average gap (dashed line). The finite subgap DoS for the solid curve is due to the normal-state puddles (note that we have not included the effect of the zero-bias anomaly in our model). The average DoS clearly displays broader features than the non-averaged one, and the outer peak positions are at higher energy.

\section{Tunneling data analysis with phenomenological Dynes broadening}
\label{app:Dynes}

Since its proposal by Dynes and collaborator in 1978~\cite{dynes}, fitting of (normalized) tunneling density of state data is often performed by using the following expression
\begin{equation}\label{eq:dynes}
\nu(\epsilon) = \frac{1}{2} \sum_{\sigma=\pm1} \mathrm{Re}\left[\frac{\epsilon-\sigma E_Z +i\gamma }{\sqrt{(\epsilon -\sigma E_Z + i\gamma)^2-\Delta^2}}\right],
\end{equation}
here generalized to allow for Zeeman splitting in the presence of parallel magnetic field, $E_Z = \mu_B H$. In this formula, $\gamma$ is the so-called Dynes broadening; in normal/superconducting bilayers with a tunnel contact (weak proximity effect), the Dynes formula correctly accounts for the finite time electrons spend in the superconducting layer before tunneling into the normal metal~\cite{hosseinkhani}. In contrast, for a good contact between two superconducting materials (strong proximity effect), a hard gap with a square root threshold replacing the square root singularity is theoretically expected. Nonetheless, we will use this formula to analyze the experimental data; assuming $\gamma$ small compared to $\Delta$, it predicts that the maxima of the outer peaks in density of states are located at
\begin{equation}\label{eq:vpk}
eV_\mathrm{pk} \simeq \pm\left(\Delta + E_Z + \gamma/\sqrt{3}\right)
\end{equation}

\begin{figure}[!bt]
\includegraphics[width=0.47\textwidth]{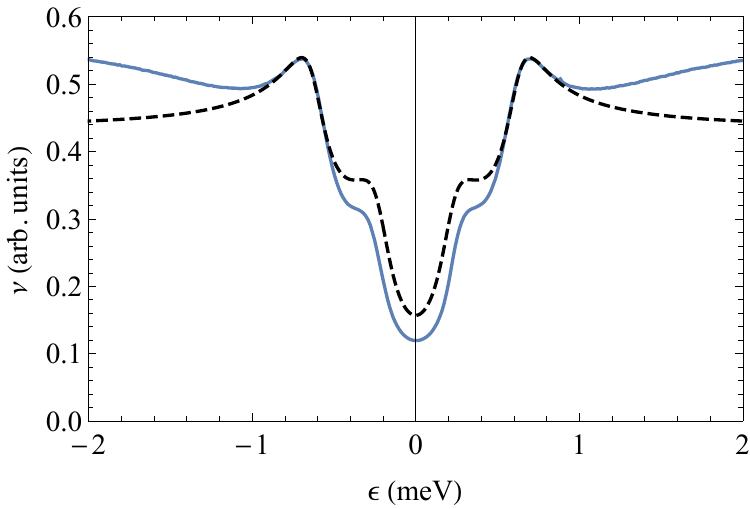}
\caption{(Color online) Blue solid line: tunneling density of states of a $d_\mathrm{Be} =0.8\,$nm Be/Al bilayer measured at 70~mK in a parallel field of 3.5~T. Black dashed line: fit to the outer peaks using the phenomenological Dynes formula, Eq.~(\ref{eq:dynes}).}
\label{fig:3.5T}
\end{figure}

In the above equations, $\Delta$ is the self-consistently determined order parameter, which is suppressed in comparison with the order parameter $\Delta_0$ that one would obtain in the absence of broadening:
\begin{equation}\label{eq:scd}
  \Delta \simeq \Delta_0\sqrt{1- 2\gamma/\Delta_0}
\end{equation}
To account phenomenologically for the orbital pair-breaking effect of the parallel magnetic field, we assume for $\gamma$ a field dependence in the form
\begin{equation}\label{eq:gez}
\gamma = \gamma_0 + c \frac{E_Z^2}{\Delta_0}
\end{equation}
Substitution of Eqs.~(\ref{eq:scd}) and (\ref{eq:gez}) into Eq.~(\ref{eq:vpk}) leads to a sublinear evolution of the peak position with parallel field.

We have used Eq.~(\ref{eq:dynes}) to fit the outer peaks of tunneling density of states data for a $d_\mathrm{Be} =0.8\,$nm bilayer measured at $70\,$mK in increasing parallel field $H$ up to 5.5~T. We treat $\gamma$ and $\Delta$ as fit parameters; as shown in Fig.~\ref{fig:3.5T}, the outer peaks can be well captured by Eq.~(\ref{eq:dynes}), and the underestimation at energies further away from the Fermi energy is due to the logarithmic behavior of the zero-bias anomaly (see also Figs.~\ref{Spectra} and \ref{BG}). The density of states at the inner peaks and near the Fermi energy is overestimated, demonstrating that at finite field the simple Dynes phenomenology cannot accurately describe the data.

\begin{figure}[!tb]
\includegraphics[width=0.47\textwidth]{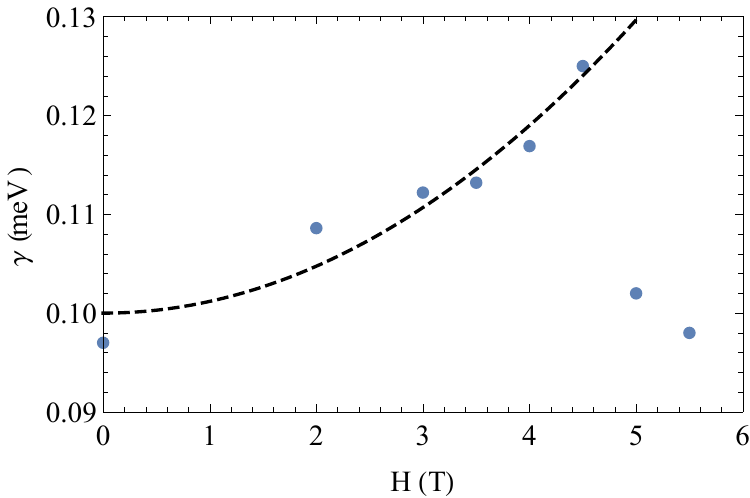}
\caption{(Color online) Dots: Dynes broadening $\gamma$ extracted by fitting the outer peaks (cf. Fig.~\ref{fig:3.5T}) at various fields . Black dashed line: fit to the data by Eq.~(\ref{eq:gez}) for fields up to 4.5~T. See text for details.}
\label{fig:gamma}
\end{figure}

\begin{figure}[!tb]
\includegraphics[width=0.47\textwidth]{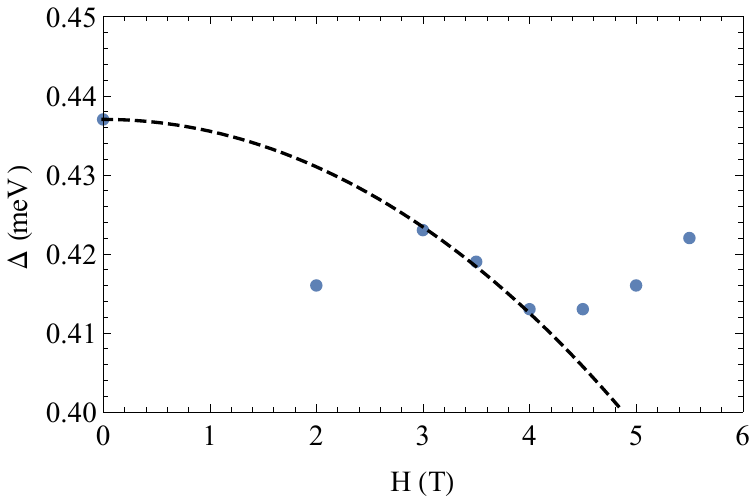}
\caption{(Color online) Dots: gap parameter $\Delta$ extracted by fitting the outer peaks (cf. Fig.~\ref{fig:3.5T}) at various fields . Black dashed line: plot of Eq.~(\ref{eq:gez}) using the parameters extracted from Fig.~\ref{fig:gamma}. See text for details.}
\label{fig:delta}
\end{figure}

In Fig.~\ref{fig:gamma} we present the extracted values of the broadening $\gamma$ as function of field. The dashed line is obtained by fitting Eq.~(\ref{eq:gez}) to the data up to 4.5~T, excluding the two data points at the highest fields. From this fit and the fitted value of $\Delta$ at zero field, we find $\gamma_0 \approx 0.1\,$meV, $c\approx0.19$, and $\Delta_0 \approx 0.55\,$meV. Interestingly, these parameters also satisfactorily describe the behavior of $\Delta$ as function of field, see Fig.~\ref{fig:delta}.
As the field increases above 4.5~T, we see that the broadening $\gamma$ decreases and the gap parameter $\Delta$ increases; this counterintuitive behavior is qualitatively in agreement with what one expects if some regions of the film with weaker superconductivity turn normal, since the surviving regions have on average a larger gap and the gap distribution of the surviving superconducting regions is narrower.

\end{document}